\documentclass[%
superscriptaddress,
 amsmath,amssymb,
 aps,
floatfix,
prapplied, twocolumn
]{revtex4-2}

\usepackage{graphicx}
\usepackage{amsmath, amssymb}
\usepackage{mathtools}
\usepackage{multirow}
\usepackage{tikz}

\def\<{\langle}
\def\>{\rangle}

\begin{document}

\title{Efficient OPA tomography of non-Gaussian states of light}
\author{Éva Rácz}
\email{racz@optics.upol.cz}
\affiliation{Department of Optics, Palacký University, 17. listopadu 1192/12, 771 46 Olomouc, Czech Republic}
\author{László Ruppert}
\affiliation{Department of Optics, Palacký University, 17. listopadu 1192/12, 771 46 Olomouc, Czech Republic}
\author{Radim Filip}
\affiliation{Department of Optics, Palacký University, 17. listopadu 1192/12, 771 46 Olomouc, Czech Republic}




\begin{abstract}
{

Current advances in nonlinear optics have made it possible to perform a homodyne-like tomography of an unknown state without highly efficient detectors or a strong local oscillator. Thereby, a new experimental direction has been opened into multimode and large-bandwidth quantum optics. An optical parametric amplifier (OPA) allows us to reconstruct the quadrature distribution of an unknown state directly from the measured intensity distribution with high precision. We propose adding a controllable displacement to the standard scheme, obtaining an improved method applicable even to asymmetric and non-Gaussian states while significantly increasing estimation accuracy and lowering the OPA amplification requirement. To demonstrate the power of our method, we accurately detect the sub-Planck phase-space structure by a distillable squeezing from the OPA estimates of various non-Gaussian states. With the improvements, OPA tomography became a generally applicable loss-tolerant and efficient alternative to OPA-assisted homodyne detection.
}
\end{abstract}


\maketitle


\section{Introduction}

Homodyne detection is a well-established phase-sensitive technique in quantum optics and quantum technology to measure nonclassical \cite{Breitenbach1997a, Welsch1999} and quantum non-Gaussian states of light \cite{Wenger2004}. Simultaneously, it is a base for quantum tomography \cite{Vogel1989, DAriano1994,  Leonhardt1996, DAriano2003} capable of reconstructing the density matrix and its equivalents like the characteristic function \cite{Lai1989, Braunstein1990}, and the Wigner function \cite{Neumann1955, Baune2017}. This technique is based on highly efficient photodetection (approaching 99.9 percent efficiency) and a highly intensive local oscillator that filters only the signal modes that match it. Due to losses, the reconstructed states may miss Wigner negativity and squeezing \cite{Braunstein1991, Breitenbach1997b}. However, non-Gaussian states may still contain nonclassical sub-Planck structures in the phase space \cite{Zurek2001}. Many times, experimental setups in quantum optics, but also other fields using the optical readout, like quantum optomechanics \cite{Aspelmeyer2014, Khalili2016}, atomic ensemble experiments \cite{Hammerer2010}, and cavity quantum electrodynamics with atoms \cite{Walther2006} have been limited by adjustments to such homodyne detection. 

Recently, to achieve phase-sensitive terahertz-order detection bandwidth \cite{Inoue2023} and measurement of signal in spectral domain \cite{RojasSantana2022}, parametric pre-amplification \cite{Knyazev2018, Takanashi2020, Shaked2018, Li2019} has been experimentally tested before the homodyne detection or even intensity detector. It has been proposed as a viable solution for realizing all-optical quantum information processing with over-THz clock frequency \cite{Budinger2022}, and high-dimensional quantum information applications \cite{Ra2020, Erhard2020}. Parametric amplification has also been shown to be useful in improving the loss-tolerance in an interferometric scheme \cite{Tian2023}. However, a systematic analysis and comparison of these estimation techniques for available quantum non-Gaussian states of light are missing, including possible techniques to lower the estimation error.

The primary interest in the experimental study by Kalash \emph{et al.}\ \cite{Kalash23} lies in the fact that by using a high-gain OPA, they could estimate quadrature distributions solely from intensity measurements. Their post-processing technique, however, can be refined and generalized to asymmetrically distributed source quadratures, as we will explain and demonstrate in the current manuscript. Furthermore, we also explore how different types of noise limit the applicability of our approach and how amending the original measurement setup to include a displacement can improve estimation accuracy. We present two approaches to estimating asymmetric quadrature probability densities: a) using a single displacement large enough to ensure the positivity of the support of the quadrature distribution b) using two smaller displacements and a consequent minimization based on the observed histogram of intensities.

In this paper, we give our model of the investigated OPA tomography scheme (including imperfections) and then describe our improved tomography methods. Then we give our results, which contain a detailed analysis of the improvement of estimation efficiency for squeezed states; we show high estimation quality for different non-Gaussian states, we compare the efficiencies of low-efficiency intensity and homodyne detectors and provide an efficient estimate of distillable squeezing \cite{Heersink2006, Filip2013} from sub-Planck structures \cite{Zurek2001}. Finally, we obtain a short conclusion and provide an outlook on future development. The details of the proposed techniques and robustness analysis against different imperfections can be found in the Appendix.


\section{Methods}\label{sec:methods}

\subsection{Model and main idea} 

\begin{figure}[t]
\centering\includegraphics[width=\columnwidth]{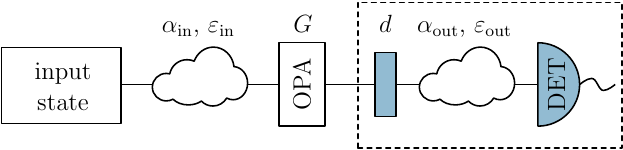}
\caption{The scheme of our model of detection in the experiments \cite{Inoue2023,Kalash23}. The state of interest is first attenuated (transmittance \(\alpha_{\mathrm{in}}\)) and is convoluted with Gaussian noise (standard deviation \(\varepsilon_{\mathrm{in}}\)). Then, it is parametrically amplified with amplification \(G\). Post-amplification, the state is displaced by \(d\) in the amplified direction. After further attenuation (\(\alpha_{\mathrm{out}}\)) and extra noise (\(\varepsilon_{\mathrm{out}}\)), the resulting signal is measured by the detector DET, which might be either an intensity detector or a homodyne detector.}\label{fig:scheme}
\end{figure}

In the following, we describe our model (see Fig.~\ref{fig:scheme}), which is based on the results in \cite{Kalash23}. 
The task is to reconstruct the probability distribution function of a quadrature using low-efficiency intensity or homodyne detection and OPA pre-amplification without making assumptions on the initial state. 
In our simulations and analysis, we will use squeezed states with and without displacement, classical non-Gaussian mixtures of two overlapping squeezed states, or quantum non-Gaussian Fock states as initial states.

The main idea behind the experimental approach is to compensate for detector inefficiency by amplifying the state in one direction using an OPA. After amplification, we measure (if not noted otherwise, we will always amplify and measure in the direction of quadrature x) either the photon number of the output state, which will dominantly contain the amplified quadrature \cite{Kalash23} or use low-efficiency and noisy homodyne detection \cite{Inoue2023}. We can manipulate the non-classical input state by a classical Gaussian operation. In the current work, we will either apply a phase-shift \(\varphi\) (to measure a different quadrature) or add a displacement d (to increase the effect of amplification further). We assume that the incoupling losses before the amplification (mainly due to imperfect visibility and optical losses) can be characterized by a loss factor \(\alpha_{\mathrm{\mathrm{in}}}\) and an added noise \(\varepsilon_{\mathrm{\mathrm{in}}}\). After the amplification, we also consider an aggregated loss factor \(\alpha_{\mathrm{out}}\) and added noise \(\varepsilon_{\mathrm{out}}\). The loss mainly comes from optical filtering, while the noise can primarily be attributed to the detector electronic noise \cite{Kalash23}.

Denoting the input quadrature variables before the amplifications by $X$ and $P$ and the squeezing parameter by $G$, the output measured by an intensity detector can be given as
\begin{equation}\label{eq}
N_x=\alpha_{\mathrm{out}} ((e^{G}\cdot X+d)^2+(e^{-G}\cdot P)^2-1/2)+\varepsilon_{\mathrm{out}}.
\end{equation}
For the sake of simplicity, we assume that the noise $\varepsilon_{\mathrm{out}}$ is normally distributed with zero mean and standard deviation $\delta_{\mathrm{out}}$. We describe the pre-amplification quadrature $X$ as a function of the initial quadrature $x_i$ as 
\begin{equation}\label{eq_init}
X=\sqrt{\alpha_{\mathrm{\mathrm{in}}}} x_i+\varepsilon_{\mathrm{\mathrm{in}}},
\end{equation}
where $\varepsilon_{\mathrm{\mathrm{in}}}$ is normally distributed with zero mean and standard deviation $\delta_{\mathrm{\mathrm{in}}}$. 
Typically, $\alpha_{\mathrm{\mathrm{in}}}$ is close to unity. 
If we apply a strong noiseless amplification $G$, then in \eqref{eq}, the first term will be significantly larger than any other, so we can approximate it with
\begin{equation}\label{approx}
N_x \approx  \alpha_{\mathrm{out}} (\alpha_{\mathrm{\mathrm{in}}} e^{G}\cdot x_i+d)^2.
\end{equation}
From this, we can express the estimate of the initial quadrature variable as
\begin{equation}\label{approx_est}
\hat{x_i} := \pm \sqrt{N_x/(e^{2G}\alpha_{\mathrm{\mathrm{in}}} \alpha_{\mathrm{out}} )} - d/(e^{G}\sqrt{\alpha_{\mathrm{\mathrm{in}}}}).
\end{equation}
But this approximation works only with some caveats: \(G\) must be high enough for \eqref{approx_est} to be applicable and, more importantly, the two solutions (\(\pm\)) in the expression are not distinguishable. Therefore, we will investigate ways to advance this basic idea and thereby significantly extend the applicability and improve the efficiency of OPA tomography. We also explore the scenario when the intensity detector is exchanged for homodyne detection \cite{Inoue2023}.


\subsection{Improved tomography method}

\begin{figure}[!t]
\begin{center}
  \includegraphics[width=0.8\columnwidth]{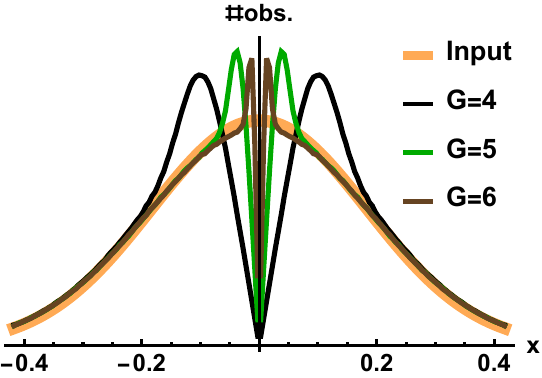}
	 \caption{The probability distribution of a Gaussian state (orange line) compared to the estimate obtained by the standard method using different parameters. The estimates are using different amplifications: $G=4$ (black line), $G=5$ (green line), $G=6$ (brown line). The precision is improved by increasing the OPA amplification; still, in the plotted typical regime of parameters, all estimators have a problem in the vicinity of zero. Simulation parameters can be found at the beginning of Sec.\ \ref{sec:photon.numbers}.\label{fig:standard}}
\end{center}
\end{figure}

The original approach in \cite{Kalash23} consisted of preparing a histogram of the measured photon numbers, and assuming that the distribution of the \(X\) quadrature is symmetric about the origin, equation \eqref{approx_est} allowed reconstructing the histogram of \(|X|\).
Albeit straightforward, there are a few problems with this method:
\begin{enumerate}
\item The binning for the reconstructed histogram is more frequent for large values while very rare for low values (due to applying the binning directly to the outcome and then transforming them during the reconstruction process).
\item The precision of the estimator is worse for low values of \(X\) (due to the relatively small amplification effect for low-energy input states, see Fig. \ref{fig:standard}).
\item The method only provides statistics of $|X|$ (thus, asymmetric input states cannot be estimated). 
\end{enumerate}

One could argue that if $G$ is large enough, that would resolve the situation. This is true to an extent (see Fig. \ref{fig:standard}) as with using large enough bins the problems around zeros might disappear (the positive and negative errors are included both in a single bin). However, it might be difficult to achieve in practice a high level of amplification without adding extra noise to the system. Also, this does not solve the issue of asymmetric input states. Moreover, many non-Gaussian states can have interesting nonclassical sub-Planck structures at the origin, as it is for amplitude or phase-diffused states \cite{Heersink2006, Franzen2006} and the Fock and cat states \cite{Lvovsky2001, Ourjoumtsev2006, Ourjoumtsev2007, Bimbard2010, Yukawa2013, Huang2015}.

To improve the method, we suggest the following changes:
\paragraph{Consistent binning} 

The first issue is that instead of making a linear binning on the measurement outcome, we first do the transformation back to the quadrature $X$ and only then define linear bins on that scale. With this, we can have a consistent binning with a fixed predefined resolution, which ensures that we will get a proper description of the whole probability distribution evenly (instead of having fewer points around zero, and more dense binning for very high values). 


\paragraph{Displacing the input}

An efficient way to avoid problems with low-intensity values is displacing the input state. It is a technique that is broadly exploited in quantum optics, even for quantum non-Gaussian states like Fock states and their superpositions \cite{Takeda2013, Miwa2014, Sakaguchi2023}. We perform the estimation for this shifted distribution, and at the end, we apply an inverse displacement to get back the original distribution. This approach circumvents the issue that the amplification is not very effective about zero (due to it being multiplicative). If we add a fixed displacement $d$ to the quadrature $X$, then the previously problematic part will be shifted out from the vicinity of zero. So the amplification will be effective again (or in other words, the ineffective amplification will be shifted around the values of $-d$). Note that with this transformation, we can also compensate for a limited value of $G$, as with the added displacement, the amplified quadrature will dominate anyway.

\paragraph{Reconstructing asymmetric distributions}


We can use the displacement of input states to reconstruct asymmetric distributions. This is possible because by applying the amplification on two different signal states, we will get twice as many data points. So by simple dimension analysis, we can obtain a reconstruction method from the estimated values of $|X|$  (see for more details in the Appendix A). 
Another important feature is that if the displacement is high enough, then the significant part of the initial distribution can be transferred completely to the positive part, which makes the estimation process much simpler.



\section{Estimation using photon number detection}\label{sec:photon.numbers}

In the following, we will investigate the estimation properties for a set of input states by generating samples of $N$ independent measurements and then applying the estimator to each sample. If not noted otherwise, we use the system parameters: $G=4$, $\delta G=0.01$, $\alpha_{\mathrm{\mathrm{in}}} = 0.99$, \(\varepsilon_{\mathrm{\mathrm{in}}} = 1 - \alpha_{\mathrm{\mathrm{in}}}\), $\alpha_{\mathrm{out}} = 0.1$, $\delta\alpha_{\mathrm{out}}=10^{-3}$, $\varepsilon_{\mathrm{out}}=3$, which corresponds to a realistic, stable setup. For histograms, we use a \(w = 0.05\) bin width and a sample size of $N=10^5$. Note that for realistic systems, $N$ could be lower: if this is the case, one should use fewer bins (larger bin width $w$), but for this manuscript, we use these numbers to obtain more detailed graphs.

We will use fidelity as a figure of merit to compare the estimated histogram with the true input distribution.
More specifically, we have to compare discretized classical probability distributions. So if we have reconstructed a histogram with relative frequency $\nu_i$ in bin \(i\), and the theoretical distribution yields $p_i$ for the same bin, then the fidelity is defined as
\begin{equation}
\label{eq:AE}
F=\Big(\sum_i\sqrt{\nu_i\cdot p_i}\Big)^2.
\end{equation}
This quantity is between zero and one, with one corresponding to perfect overlap. Therefore, the infidelity \(1-F\) can be considered as the error of the estimation. 
The probabilities \(p_i\) can be determined using the cumulative distribution function \(H\) of the input distribution:
\[p_i = H(x_i+w/2) - H(x_i - w/2),\]
with \(x_i\) denoting the center of bin \(i\).
If \(w\) is small enough, of course, \(p_i \approx h(x_i) \cdot w\), with \(h\) denoting the probability density function of the input.

\subsection{Estimation of squeezed states} 

\begin{figure}[!t]
\begin{center}
\includegraphics[width=0.48\columnwidth]{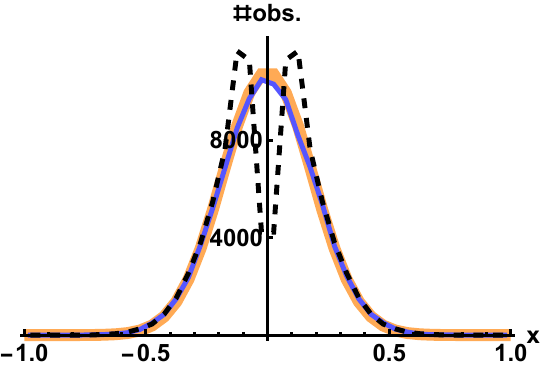}
\includegraphics[width=0.51\columnwidth]{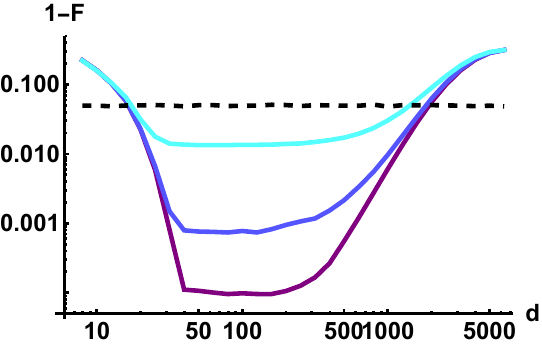}
\\
\includegraphics[width=0.48\columnwidth]{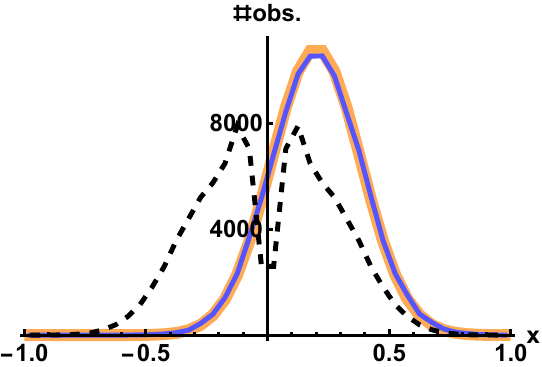}
\includegraphics[width=0.51\columnwidth]{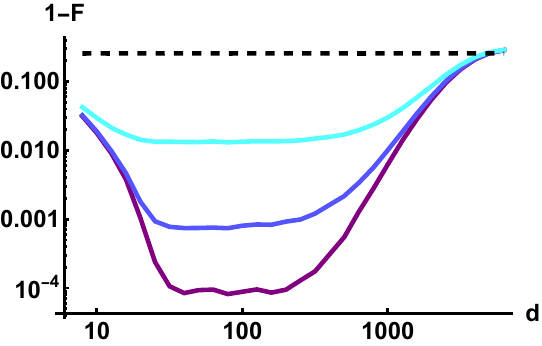}
\caption{Estimation of (top) a non-displaced squeezed state and (bottom) a displaced squeezed state (squeezing parameter \(g= 1\)). Left: Comparison of reconstructed histograms (dashed black line: original method, blue solid line: displaced approach) with the input distribution (orange line). Right: Infidelity $1-F$ as a function of the displacement $d$. The dashed black line corresponds to the standard method with \(\alpha_{\mathrm{\mathrm{in}}} = 0.95\); the light blue, blue, and purple lines to the improved method with \(\alpha_{\mathrm{\mathrm{in}}} = 0.95\), 0.99, and 1, respectively. \label{fig:sq}}
\end{center}
\end{figure}

First, let us check the estimation of a non-displaced squeezed state (Fig.~\ref{fig:sq} top left). We can see that even with imperfections present, we can estimate the original distribution with high precision using our method. The standard method has similar properties, except for some problems around zero, where signal amplification is not effective. 

If we compare the efficiencies of these two methods (Fig.~\ref{fig:sq} top right), we can see that our method provides an improvement of about three orders of magnitude in \(1-F\) for the ideal ($\alpha_{\mathrm{in}} = 1$) setup, and an about fivefold improvement for the noisy ($\alpha_{\mathrm{in}} = 95\%$) setup. Note that this is the ideal input for the standard method, as it is a symmetric state and is not strongly squeezed (the majority of the distribution is far from zero). 

That is, adding a displacement can significantly improve the quality of our estimation. This improvement is present for a wide range of displacement values, however, we can lose the advantage if the displacement becomes too large. Heuristically, this comes from the fact that in \eqref{approx}, the term $d$ will dominate over $x_i$, so any fluctuation in the transmittances and the squeezing parameter will be multiplied by that large value. 



If we displace the input squeezed state, the situation remains similar (Fig.\ \ref{fig:sq} bottom left). The main difference is that the standard method becomes even less efficient. This is not surprising, as even a small displacement will break the needed symmetry for that method, i.e., the asymmetry causes further errors (see Fig.\ \ref{fig:sq} bottom left). In contrast, our method is not sensitive to displacement; therefore, we can obtain similar efficiencies as in the zero-displacement case. More precisely, the displacement in the input state acts similarly to our additional displacement, so the original displacement will increase the efficiency for low values of the displacement $d$.

\subsection{Estimation of non-Gaussian states} 

In the following, we will investigate the case of the mixing of squeezed states. This is a non-Gaussian state, so for example, we cannot reconstruct the distribution of the input from its variance; therefore, a full tomography of the distribution is necessary in all cases. 

\begin{figure}[!t]
\begin{center}
  \includegraphics[width=0.48\columnwidth]{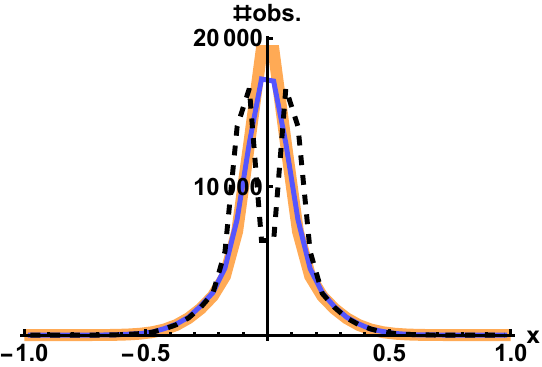}
	\includegraphics[width=0.51\columnwidth]{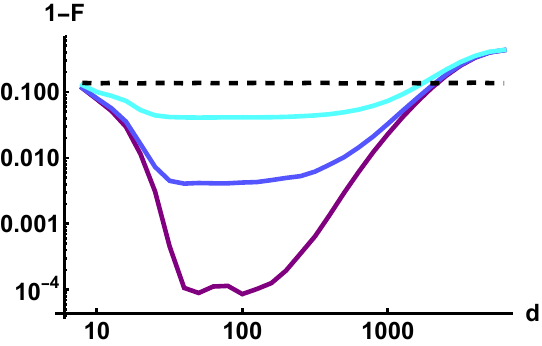}
 \\
   \includegraphics[width=0.48\columnwidth]{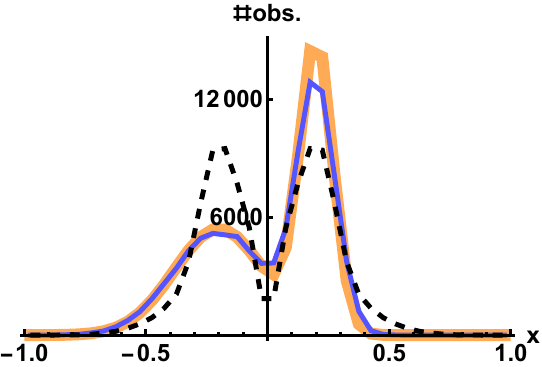}
	\includegraphics[width=0.51\columnwidth]{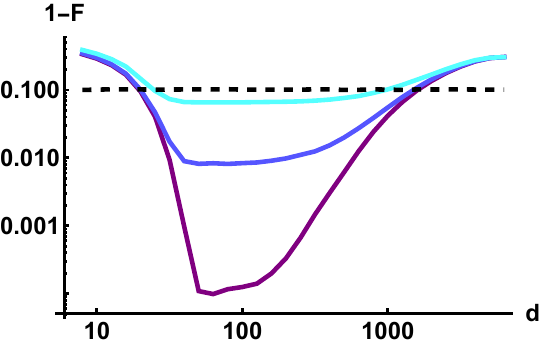}
\caption{Estimation of (top) a symmetric non-Gaussian state: a 50\%-50\% mixture of two non-displaced squeezed states ($g_A=2$ and $g_B=1$) and (bottom) an asymmetric non-Gaussian state the same mixture with displacement values of $0.2$ and $-0.2$, respectively. Left: Comparison of reconstructed histograms (dashed black line: original method, blue solid line: displaced approach) with the input distribution (orange line). Right: Infidelity $1-F$ as a function of the displacement $d$. The dashed black line corresponds to the standard method with \(\alpha_{\mathrm{\mathrm{in}}} = 0.95\); the light blue, blue, and purple lines to the improved method with \(\alpha_{\mathrm{\mathrm{in}}} = 0.95\), 0.99, and 1, respectively.  \label{fig:2sq}}
\end{center}
\end{figure}

If we check the histogram of the non-displaced case (Fig.~ \ref{fig:2sq}, top left), the main difference compared to the simple squeezed state \#1 is that the more squeezed ($g_A=2$) counterpart adds a larger peak about zero. Again, this reduces the efficiency of the standard method (Fig.~\ref{fig:2sq}, top right dashed black lines) since there are more observations close to zero, which are strongly affected by noise. Our method behaves similarly as for the squeezed state (Fig.~\ref{fig:sq}): the errors are a bit larger, especially for the noisier case, since adding noise to a more squeezed state produces a larger change in its distribution. 

  

We also investigate the case of the mixture of displaced squeezed states (Fig.~ \ref{fig:2sq}, bottom left). This distribution is problematic for the standard method both regarding asymmetry and the inefficiency around zero. Despite these issues, the figure of merit is similar to the previous, symmetric case. Even so, its efficiency will only be comparable to our method for the case of noisy input states ($\alpha_{\mathrm{in}}=95\%$), for the ideal and less noisy cases, our method still significantly outperforms the original approach. 

\subsection{Estimation of Fock states} 

Finally, we examine different Fock states. These are symmetric non-Gaussian states which have a negative Wigner function. For the sake of simplicity and better comparison, we will again estimate the marginal distribution of the quadrature $x$, which is a classical probability distribution.

\begin{figure}[!t]
\begin{center}
  \includegraphics[width=0.48\columnwidth]{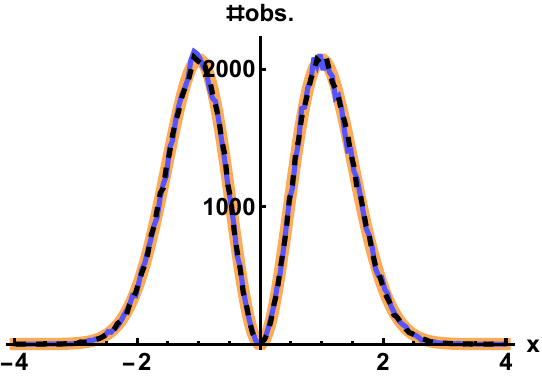}
	\includegraphics[width=0.51\columnwidth]{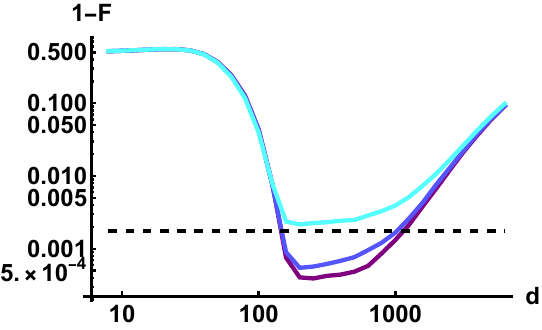}
 \\
  \includegraphics[width=0.48\columnwidth]{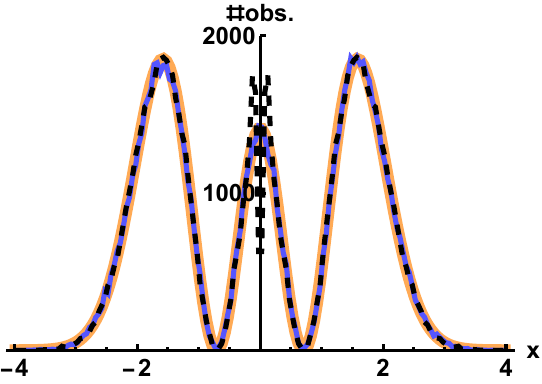}
	\includegraphics[width=0.51\columnwidth]{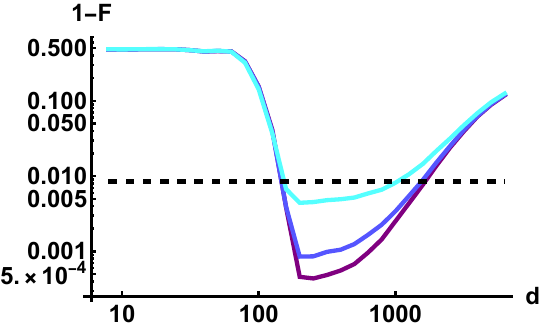}
\caption{Estimation of (top) the single-photon state and (bottom)  the two-photon state. Left: Comparison of reconstructed histograms (dashed black line: original method, blue solid line: displaced approach) with the input distribution (orange line). Right: $1-F$ as a function of the displacement $d$. The dashed black line corresponds to the standard method with \(\alpha_{\mathrm{\mathrm{in}}} = 0.95\); the light blue, blue, and purple lines to the improved method with \(\alpha_{\mathrm{\mathrm{in}}} = 0.95\), 0.99, and 1, respectively. \label{fig:Fock}}
\end{center}
\end{figure}

By checking the histogram of the single-photon case (Fig.~\ref{fig:Fock}, top left), we can see that this is the only case that could be estimated very efficiently even with the standard method. The reason is simple: this state is symmetric and its density around zero is negligible, so the observed errors around very low photon numbers are insignificant. This high efficiency of the standard method is also apparent in Fig.~\ref{fig:Fock} (top right, black dashed line). We can obtain only a small improvement and only for a limited range of displacement values (close to ideal). 
Therefore, for reconstructing the distribution of a single-photon state (or a similar symmetric state with a low density about the origin) the original method is sufficient. 
  

Examining the two-photon state (Fig.~\ref{fig:Fock}, bottom left), we see a performance that is between the single-photon and the squeezed states. Once again, the issue with the standard method around zero is quite visible, but since the distribution is not so focused around zero as for symmetric squeezed states, the relative magnitude of the error is smaller. So compared to the previous case, the fidelity of the standard method is lower (Fig.~\ref{fig:Fock}, bottom right black dashed line); therefore, our method provides a more significant improvement over it. But as in the previous symmetric cases. 
Note also that the obtained efficiency is very high for either case: we can estimate Fock states much more precisely than the different squeezed states. The reason for this is simple: the distribution of these states is less focused around the zero, so we amplify much larger values, and the estimation is more efficient for larger photon numbers. 

\section{Dependence on amplification}

As suggested earlier, one of the key factors in estimation efficiency is the amplification strength $G$, so in the following, we will investigate its effect.

\begin{figure}[!htbp]
\begin{center}
  \includegraphics[width=0.8\columnwidth]{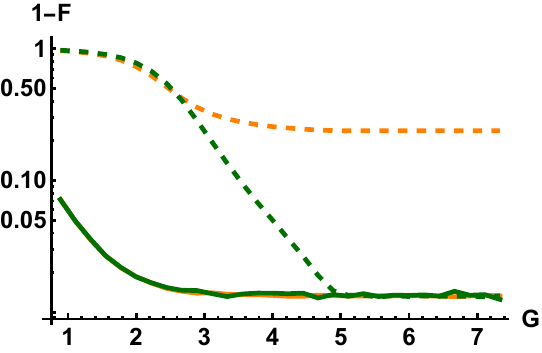}
 \caption{Infidelity of the protocol as a function of the amplification. Green lines: input is a non-displaced squeezed state (Fig.~\ref{fig:sq} top); yellow lines: input is a displaced squeezed state (Fig.~\ref{fig:sq} bottom). Dashed lines: original method; solid lines: our method. \label{fig:G_squeezed}}
\end{center}
\end{figure}

Firstly, for non-displaced squeezed state inputs (Fig.\ \ref{fig:G_squeezed}, green lines), then we observe that both our method (solid) and the standard method (dashed) saturate at the same low level of error. However, there is a significant difference in when those methods achieve these saturated values: the standard method needs at least $G=5$, while our method achieves the same performance already at $G\approx3$. So using our method, much lower amplification suffices to achieve optimal performance, which might be useful in practical applications.

On the other hand, the difference is even more significant for the displaced squeezed state (Fig.~\ref{fig:G_squeezed}, yellow lines). Our method (solid) has the same efficiency as before, as the method is not sensitive to displacement. In contrast, the standard method cannot handle asymmetric states, so it saturates at a significantly lower level of fidelity, making the estimation completely wrong (see also Fig.~\ref{fig:sq} bottom left)).

\begin{figure}[!t]
\begin{center}
  \includegraphics[width=0.8\columnwidth]{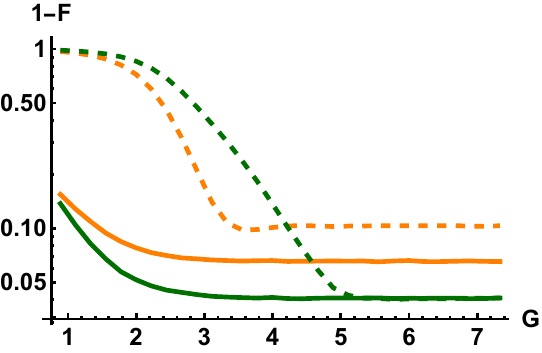}
 \caption{Infidelity of the protocol as a function of the amplification. Green lines: input is a mixture of non-displaced squeezed states (Fig.~\ref{fig:2sq} top); yellow lines: input is a mixture of displaced squeezed states (Fig.~\ref{fig:2sq} bottom). Dashed lines: original method; solid lines: our method.\label{fig:G_mixture}}
\end{center}
\end{figure}

We can come to similar conclusions considering mixtures of squeezed states (Fig.\ \ref{fig:G_mixture}). For the symmetric scenario (green lines), the estimates have exactly the same properties as before. The only difference is that the saturated level of error is a little higher (0.04 vs 0.02), which is caused by the bigger bias of the larger squeezing (adding a fixed amount of noise to a larger squeezing causes a larger change in the probability distribution). In the asymmetric case, where both constituents in the mixture are displaced (Fig.~\ref{fig:2sq}, bottom right), the fidelity of either method decreases. But again, our method saturates already at a low level of amplification ($G\approx 3$). This is remarkable as one of the peaks in the distribution has a squeezing factor of $g_A=2$, and we use a similar strength of amplification to achieve optimal performance. Again, the standard method saturates at a higher level as it cannot handle asymmetric distributions. 

\begin{figure}[!b]
\begin{center}
  \includegraphics[width=0.8\columnwidth]{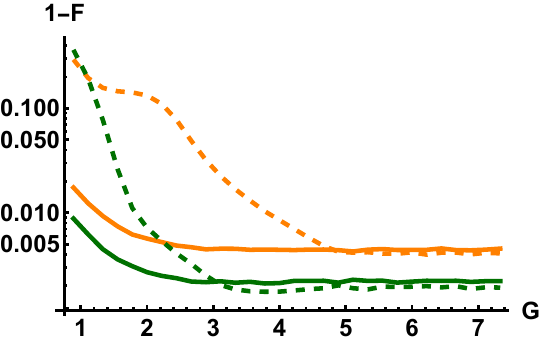}
 \caption{Infidelity of the protocol as a function of the amplification. Green lines: input a single-photon state (Fig.~\ref{fig:Fock} top); yellow lines: input is a two-photon state (Fig.~\ref{fig:Fock} bottom). Dashed lines: original method; solid lines: our method.\label{fig:G_Fock}}
\end{center}
\end{figure}

Finally, for Fock states (Fig.\ \ref{fig:G_Fock}), the situation is similar to the other symmetric cases: our method and the standard method converge to the same values in both the Fock-1 (green) and Fock-2 (yellow) cases. Again our method works better for lower values of $G$, which is more prominent for the Fock-2 state. The obtained optimal fidelity is a bit lower for the Fock-2 state (which is not surprising), but even there, the estimation efficiency is better than in all previous squeezed cases. 

Of course, the efficiency of the estimation depends on the other parameters of the system as well. We checked the robustness of our method (more details in the Appendix B), and we can conclude that if all the different errors are below a threshold value, we can achieve a similar magnitude of errors and the proposed method efficiently reconstructs the unknown state. While above that threshold, the errors start to increase rapidly. 

The pre-amplification noise ($\varepsilon_{\mathrm{in}}$) is the most significant of the different types of noises. This is because, in realistic scenarios, that is the only case where we cannot easily get below the critical threshold. This is problematic because the added noise is amplified together with the signal, so this problem cannot be easily circumvented by other improvements (e.g., by increasing the value of $G$). So if the value of $\varepsilon_{\mathrm{in}}$ is large, we have a significant bias in our estimation. Knowing its value, we can try to deconvolute the added Gaussian noise from our estimated histogram, but this is usually a difficult task, and mostly works well for states that are close to Gaussian.

\section{Comparison with homodyne detection} 

In the following, we will compare our obtained results with the case of homodyne detection, i.e., when in 
Fig.\ \ref{fig:scheme}, the detector part is not a simple photon number detector, but we use a local oscillator to obtain the quadrature in a specific direction. If we again do the amplification in direction x, we can observe the photon influx difference:
\begin{equation}\label{eq_homo}
i_x=(\sqrt{\alpha_{out}} (e^{G}\cdot X+d)+\varepsilon_{out}) \cdot \alpha_{LO}.
\end{equation}
where $\alpha_{LO}$ is the strength of local oscillator, and as before, we have
\begin{equation}\label{eq_init2}
X=\sqrt{\alpha_{\mathrm{\mathrm{in}}}} x_i+\varepsilon_{\mathrm{\mathrm{in}}}.
\end{equation}

The post-amplification loss ($\alpha_{out}$) and noise ($\varepsilon_{out}$) can consist of multiple sources. For the sake of simplification we will only use in our simulation the dominant parts: we will assume that loss is caused by the detection efficiency ($\alpha_{out}=\eta_{LO}$), while the noise is coming from the noise added due to imperfect detection efficiency and electronic noise ($\varepsilon_{out}=\sqrt{1-\eta_{LO}} \cdot \varepsilon_{0} + \varepsilon_{elec}).$

For the case of homodyne detection \eqref{eq_homo}, the situation is much simpler as the output is a linear function of the input state (instead of quadratic). In that case, we can simply obtain an unbiased estimator:
\begin{equation}\label{approx_est2}
\hat{x_i} := i_x/(e^{G} \alpha_{LO} \sqrt{\alpha_{\mathrm{\mathrm{in}}} \alpha_{out}} ) - d/(e^{G}\sqrt{\alpha_{\mathrm{\mathrm{in}}}}).
\end{equation}

\begin{figure}[!t]
\begin{center}
  \includegraphics[width=0.8\columnwidth]{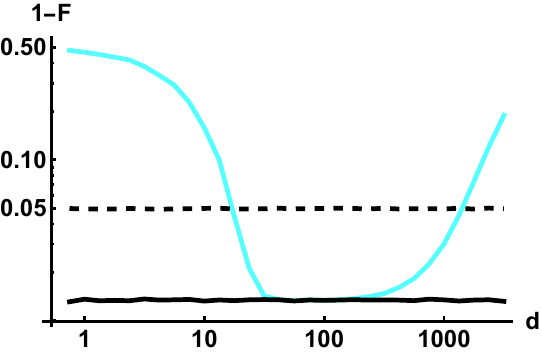}
	 \caption{Estimation of a non-displaced squeezed state (see Fig.~\ref{fig:sq} for further details), infidelity $1-F$ as a function of the displacement $d$. The dashed black line corresponds to the standard method; the light blue corresponds to the improved method, and the solid black line corresponds to the local oscillator estimation (homodyning using $\alpha_{out}=0.5$, $\varepsilon_{elec}=0.1$). \label{fig:LO1}}
\end{center}
\end{figure}

If we compare this protocol with our method (Fig.\ \ref{fig:LO1}), we can see that in the case of homodyne detection, the efficiency does not depend on the displacement $d$. This is not surprising as the output is linear (\eqref{eq_homo}), so applying a displacement and then removing a displacement does not change anything. In the case of a photon number counting detector, the displacement was needed for the insensitivity of the scheme around zero, which is simply not an issue for homodyning. The other finding is that the efficiency of homodyne estimation is the same as our estimator for the ideal regime of displacement (where the $d$ is not too small or large). 

So, in general, we can conclude that the homodyne estimation is mathematically a simpler solution to achieve the ideal estimation efficiency, but with a proper displacement, we can achieve the same efficiency with a single photon counter detector. The basic idea between the two estimation methods is similar, thus, not so surprisingly, the same performance. So the applicability of the different detectors mainly depends on the physical specification of the setup, e.g., for a heavily multimode source the phase matching with the local oscillator could be more problematic.

\begin{figure}[!t]
\begin{center}
  \includegraphics[width=0.8\columnwidth]{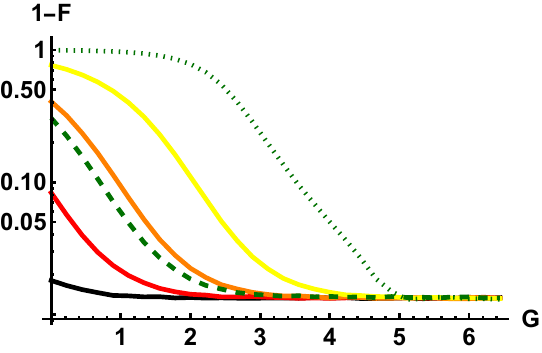}
	 \caption{Infidelity of the homodyne detection protocol as a function of the OPA amplification $G$. We plotted lines for $\alpha_{out}=1$ (black),  $\alpha_{out}=0.9$ (red),  $\alpha_{out}=0.5$ (orange), and  $\alpha_{out}=0.1$ (yellow). For comparison, we plotted the standard OPA method (green dotted) and our improved OPA method (green dashed) for $\alpha_{out}=0.1$. We have $\varepsilon_{elec}=0.1$, otherwise the same parameters as in  Fig.~\ref{fig:sq}.\label{fig:LO2}}
\end{center}
\end{figure}

If we compare the homodyne estimation with different detector efficiencies  (Fig.\ \ref{fig:LO2}), then we can conclude that they act very similarly. If the OPA gain $G$ is large enough, then we could saturate the optimal performance. The place of this saturation is determined by the parameters of the system, mainly the post-amplification noise $\varepsilon_{out}$. But we could achieve even for large noise ($\varepsilon=0.1$) the ideal performance for relatively low amplification ($G\approx 4$). Below this threshold, the estimation efficiency decreases. This decrease in performance is faster for lower detector efficiency, but if we are not in the regime of very low detector efficiencies, then we could get similar efficiencies. 

Comparing this result to our improved OPA tomography method (green dashed line), we can see that the efficiency will be similar for the homodyne detection with $\alpha_{out}=0.5$. But we received this estimation performance using $\alpha_{out}=0.1$, thus using a much less efficient detector. Of course, the efficiencies of detectors for these two cases are not directly comparable as although the two setups are mathematically similar, technically there are immense differences in whether someone is using a local oscillator or not. Nevertheless, this suggests that if for some technical reasons, we are in a regime of low detection efficiencies, using OPA tomography could be advantageous as it is less sensitive to losses.

\section{Distillable squeezing}

We will demonstrate the power of this newly proposed method by using the resulting histograms to uncover sub-Planck structures \cite{Zurek2001} by  distillable squeezing \cite{Filip2013}. For a Gaussian state, the level of squeezing can be determined from the value of the variance in the squeezed direction. For non-Gaussian states, one can distill squeezing from the maximal curvature of the probability density function \cite{Filip2013}. For Gaussian states, the two definitions are, of course, equivalent. However, the latter can be applied to our other investigated non-Gaussian states as well. 

Note that due to its local nature, the second definition requires the estimated histogram to be quite precise, which is why the standard method is not applicable due to its inaccuracy in the vicinity of zero (and its restriction to symmetrical states). In other words, if the maximum under consideration happens to be in zero, distilling squeezing becomes impractical.

We used a very simple method to obtain the maximal curvature: we checked our histogram for local maxima and estimated the curvature in each. We calculated the curvature for any given maximum by fitting a parabola to a certain number of bins $m$ around the maximum (the points were selected symmetrically). If the fitted parabola has a formula of $p(x)=-a (x-b)^2+c$, then according to \cite{Filip2013}, the distillable variance will be $$V_d=p(b)/|2p''(b)|=c/2a.$$ Naturally, as a parabola has three parameters, this fit requires at least three points (including the bin with maximal value). If we increase the number of bins, we will get a wider interval where we perform the fit (including more data points).

\begin{figure}[!t]
\begin{center}
  \includegraphics[width=0.49\columnwidth]{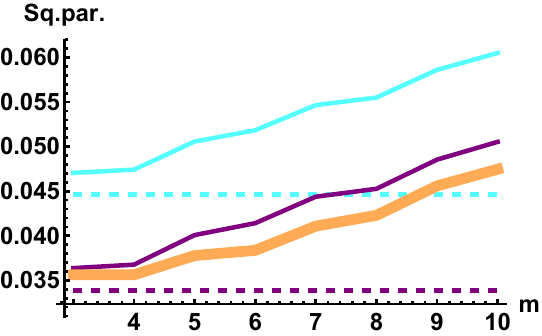}
	\includegraphics[width=0.49\columnwidth]{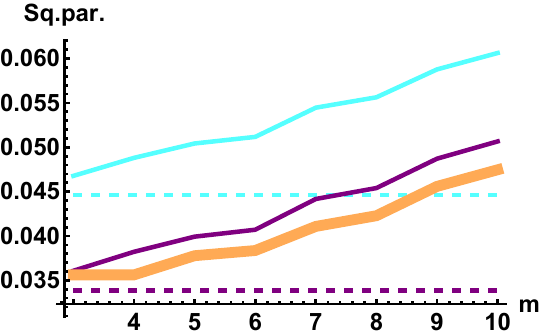}
	\includegraphics[width=0.49\columnwidth]{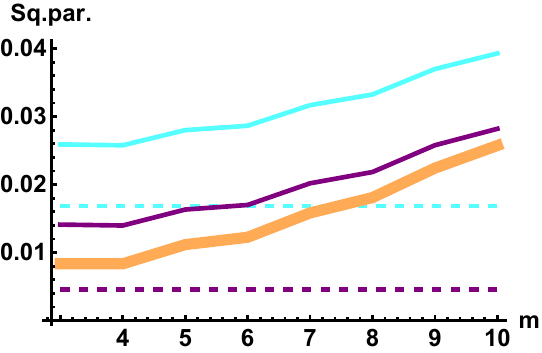}
	\includegraphics[width=0.49\columnwidth]{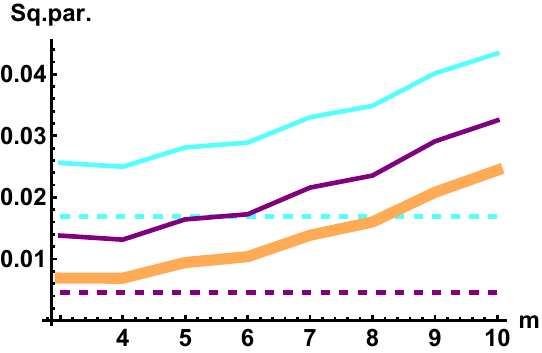}
    \includegraphics[width=0.49\columnwidth]{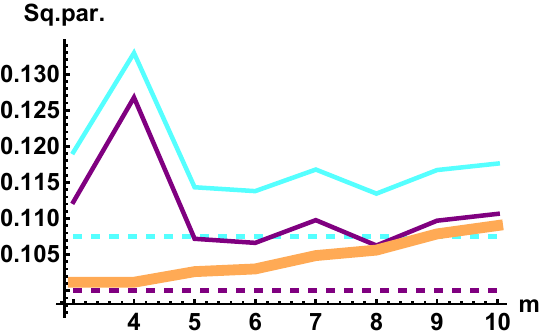}
	\includegraphics[width=0.49\columnwidth]{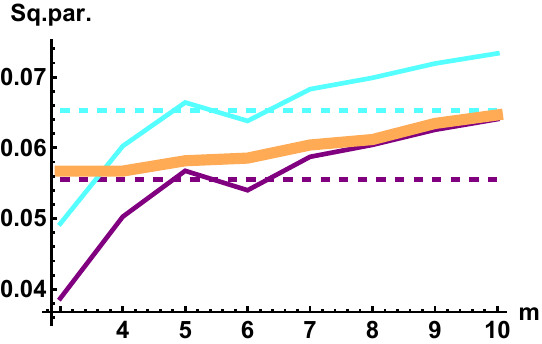}
 \caption{Distillable squeezing for different squeezed and non-Gaussian states using our improved OPA tomography scheme. The distillable squeezing parameters for different numbers of fitted bins $m$ are shown with solid lines, in comparison, the true values of the squeezing parameter are shown with dashed line. The realistic setup ($\alpha_{in}=0.95$) is plotted with light blue, the lossless/renormalized setup ($\alpha_{in}=1$) is plotted with purple, the ideal theoretical limit of the method is plotted with thick yellow. Top row: squeezed states (Fig.~\ref{fig:sq}). Middle row: mixtures of squeezed states (Fig.~\ref{fig:2sq}). Bottom row: Fock-2 (Fig.~\ref{fig:Fock}) and, for comparison, Fock-4 states. \label{fig:squeezing}}
\end{center}
\end{figure}

Figure \ref{fig:squeezing} shows the distillable squeezing from our estimated histograms. We know that knowing the value of the input loss ($\alpha_{\mathrm{in}}$), we can calculate the level of squeezing after adding this noise blue vs.\ purple dashed lines) due to the additivity of the variances. Note that removing the added Gaussian noise from the histogram is not this straightforward (it would require deconvolution). Conversely, using this equivalence, if we have a precise value for the distillable squeezing for the lossy case (blue solid lines), then from that, we can simply obtain a precise estimate of the actual value of the squeezing (purple solid lines). The thick yellow lines show the best possible values calculated from the noiseless theoretical histograms.

Ideally, our method (solid lines) should be close to the actual values of squeezing (dashed line). Our method to estimate distillable squeezing provides a slightly worse level of squeezing, but in all cases, the difference is small, meaning that we can estimate the squeezing relatively precisely. One can see that in all cases, if we use a higher number of bins $m$, our estimation becomes worse. This worsening happens because the curvature is a local property, so adding more points decreases the estimation's accuracy. However, according to our experience, adding 1-2 extra data points can help negate the statistical fluctuations of individual data points, compensating for the negative effect of including farther-off points in the estimation.

If we check the estimation of the squeezing parameter for squeezed states (top row), we can see no significant difference between the displaced and non-displaced cases. This is expected as our method is not sensitive to displacement: one only has to find the maximum of the histogram and fit a parabola. A similar conclusion is true for the mixture of squeezed states (middle row), where, unsurprisingly, our estimate is less accurate. But we could clearly see that our estimated levels of squeezing are lower than in the previous case, so with this method, we can estimate the more squeezed ($g_A=2$) part of the distribution (which would not be possible if we use the variance, which is determined mainly by the less squeezed $g_B=1$). Also, it is interesting that if the two distributions are separated by displacements (middle right), the accuracy does not improve compared to the case of overlapping non-displaced distributions (middle left). 

Finally, we investigated the distillable squeezing for two-photon and four-photon Fock states (bottom row). Note that in this case, using the method described in \cite{Filip2013}, we can approximate a squeezed state around zero using multiple copies of the given Fock states. The variance of the distribution is not useful for this purpose at all, so in this case, the only viable method to estimate the squeezing is to use the fitted curvature based on the estimated distribution. We can see that the obtained squeezing is less stable numerically than for Gaussian states. This is especially true for low values $m$. So here, including more bins $m$ for the fit could be actually useful, as it worsens the fit only a bit, but in exchange, we will get a much numerically stable fit (left vs right part of blue solid lines). Note that in this case, our estimation is closer to the real values of squeezing. This is not surprising, as in the previous subsections, we concluded that we can estimate Fock states with higher precision; thus, the distillable squeezing is also closer to the real value.

\section{Conclusion and Discussion}

We proposed an improved method for OPA tomography applicable directly to non-Gaussian states of light. The improvement's essential element is adding a displacement step post-amplification to the standard scheme proposed in \cite{Kalash23}. The displacement significantly increases the accuracy around zero, as without it, there is a ``blind spot'' that distorts the whole distribution. This effect is quite apparent for the distillation of squeezing, where--if the origin is the most likely value of the investigated quadrature--the histogram yielded by the original method becomes useless. 

Compared to the standard method, we can significantly improve the estimation efficiency given a fixed level of OPA amplification $G$ (Figs.\ \ref{fig:sq}-\ref{fig:Fock}). More importantly, our method does not assume that the input state is symmetric, so the range of applications is much broader. Furthermore, even for states symmetric about the origin, we could obtain an order of magnitude lower average error using fewer assumptions. Note that this improvement is attainable using a relatively small displacement (actually, there is a range of displacements, on the order of $d\sim 100$, where the performance is optimal; efficiency decreases for too large and too small displacements). Fock states present a special case of symmetric quantum non-Gaussian states. For Fock states, the improved method is even more precise than for Gaussian states. Note that for odd Fock states, this precision is also achievable by the standard approach since the quadrature distribution has a zero in the origin for such states.

The efficiency of our improved method saturates at a relatively low level of OPA amplification (Figs.\ \ref{fig:G_squeezed}-\ref{fig:G_Fock}), i.e., it is not necessary to have a very strong OPA (which could be challenging to achieve without increasing the noise level) to perform an efficient estimation. We also investigated the effects of different imperfections and concluded that our method is relatively robust against them. Comparing our method to homodyne detection with a pre-amplification step (that is, exchanging the photodetector for homodyning in the original scheme), we see that they saturate at the same level for increasing amplification. For our OPA tomography, we need to apply a displacement in the optimal regime to achieve this optimal performance (Fig.\ \ref{fig:LO1}). However, in exchange, a lower detection efficiency is needed to achieve the same performance (Fig.\ \ref{fig:LO2}). Finally, we investigated the precision of the estimation of distillable squeezing for different Gaussian and non-Gaussian states  (Fig.\ \ref{fig:squeezing}). We could use our high-precision OPA histograms to estimate the curvature of the probability density function about zero, which would not be feasible with the standard method.

An important note is that in our scheme (Fig.\ \ref{fig:scheme}), we used the displacement as part of the measurement setup, i.e., post-amplification. This was done because usually, in a heavily multimode scenario, applying a displacement causes significant loss, which can be easily dealt with if the loss happens after the amplification. However, if the displacement could be performed close to ideally (without causing much loss or noise), then it could be performed instead before the amplification, meaning that a minimal level of displacement could achieve optimal performance. Also, note that even though applying a displacement is fundamentally a similar concept to homodyning, technically, it is still very different as we apply orders of magnitude weaker displacement compared to the case of a local oscillator.

Overall, we achieved a significant improvement in both the applicability and the performance of OPA tomography. Our improved tomography method is feasible with a general input state (it could be asymmetrical, non-Gaussian, or both), a lower strength of OPA amplification is necessary for optimal performance, and even with having more losses, it is comparable with its alternative of using homodyne detection. Our results will inspire the field, and OPA amplification will prove to be an excellent alternative to classical estimation approaches in many different scenarios involving multimode and large telecom bandwidth states of light. Moreover, such a method can find applications to detect and identify quantum non-Gaussian states of other mechanical and atomic systems, like single phonon states quantum optomechanics \cite{Riedinger2016, Hong2017} and optically controlled atomic ensembles \cite{Wang2019, Cao2020}.





\begin{acknowledgments}
 As members of project SPARQL, R.~F., L.~R. and É.~R. acknowledge funding from the MEYS of the Czech Republic (Grant Agreement 8C22001). Project SPARQL has received funding from the European Union’s Horizon 2020 Research and Innovation Programme under Grant Agreement no. 731473 and 101017733 (QuantERA). L.~R., É.~R. and R.~F. have further been supported by the European Union’s 2020 research and innovation programme (CSA - Coordination and support action, H2020-WIDESPREAD-2020-5) under grant agreement No. 951737 (NONGAUSS). R.~F. received additional support from the grant 23-06308S of the Czech Science Foundation. The authors are thankful to Mahmoud Kalash and Maria V.~Chekhova for helpful discussions.
\end{acknowledgments}




%
\bibliography{OPA_tomography.bib}




\appendix
\onecolumngrid
\vspace{0.5cm}

\section{Detailed description of our improved method}

As stated in the main text, we can estimate an underlying PDF \(f(x)\) with high precision, even if we do not assume that it is symmetric about the origin. We propose a scheme that is composed of two different approaches: a simple one and a more complicated which relies on a linear least squares problem (we omit here all the imperfections to focus on the main points of algorithm).

To determine which method should be used, we suggest applying a displacement $d$ to the input and then one should investigate the output distribution around zero. If there are not many outcomes near the zero, that means that we applied a large enough displacement so that we move the whole (or almost the whole) distribution to the positive half-line. In this case, taking simply the square root of the output will give directly a simple estimator of the underlying displaced distribution. And so by subtracting $d$ from this distribution, we get back a simple estimation of the original input distribution.

\begin{figure}[!b]
\begin{center}
  \includegraphics[width=0.45\columnwidth]{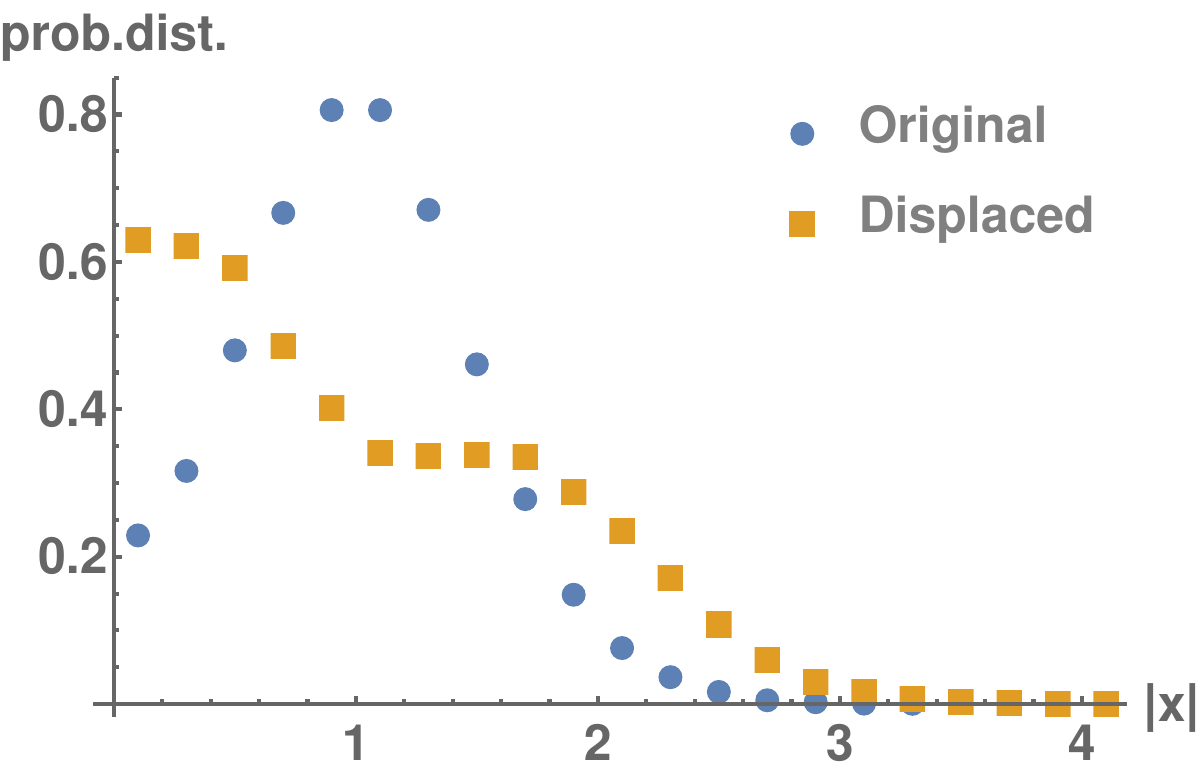}
	\includegraphics[width=0.45\columnwidth]{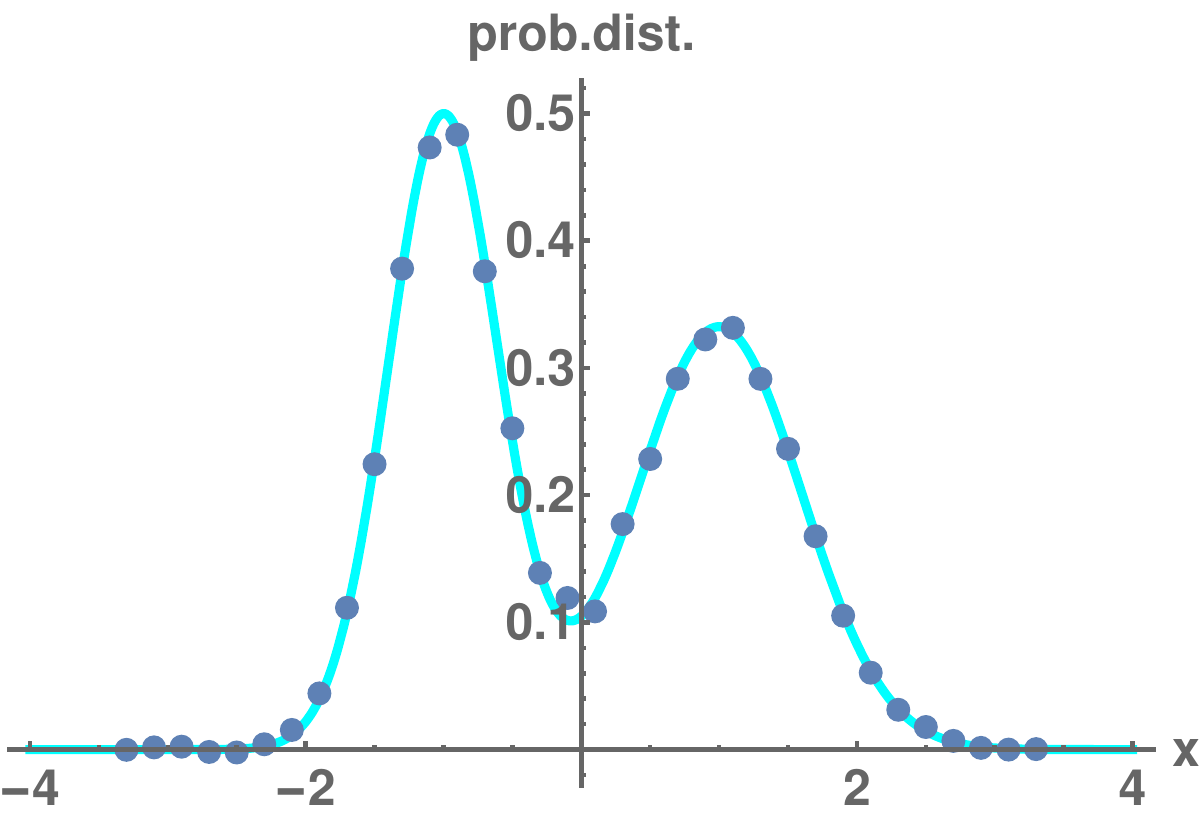}
 \caption{Reconstruction of asymmetric probability distributions. (left) Histogram of the absolute values (the negative and positive values cannot be discriminated) for the original and displaced distribution, (right) The reconstructed histogram (dark blue points) with the baseline of original distribution (light blue line). The underlying distribution was a 50-50 mixture of two Gaussians, the first centered at -1, with a standard deviation of 0.4, the second centered at 1, with a standard deviation of 0.6. The sample sizes were \(n = 10^5\), the value of the displacement was \(d = 0.6\), the resolution used for binning was \(w = 0.2\)\label{fig:asym}}
\end{center}
\end{figure}

If there are a significant number of points around zero (see for example Fig. \ref{fig:asym}, left subfigure), then we use a more involved double-modulation method detailed below:

First, we perform two measurements, with \(N/2\) observations each:
	\begin{itemize}
	\item non-displaced sample: \(\{y_1, y_2,\ldots, y_{N/2}\}\); \(y_i = \sqrt{x_i^2}\); 
 ~ the PDF of \(Y\) is \(g(y)=f(y)+f(-y)\);
	\item displaced sample: \(\{z_1, z_2,\ldots, z_{N/2}\}\); \(z_i = \sqrt{(x_i'+d)^2}\); 
 ~ the PDF of \(Z\) is \(h(y)=f(y-d)+f(-y-d)\).
	\end{itemize}

Then, we calculate the histograms of the two samples using the same binning, with bin centers in \(\frac w 2, \frac 32  w, \ldots, \left(n-\frac 1 2\right) w\); with \(n_1\) denoting the last non-empty bin for the \(y\)-sample, and \(n_2\) denoting the last non-empty bin in the \(z\)-sample:
	\begin{itemize}
		\item \(\mathbf g \equiv (g_1, g_2,\ldots, g_{n_1})^\top\)
		\item \(\mathbf h \equiv (h_1, h_2, \ldots, h_{n_2})^\top\)
	\end{itemize}
	These histograms are shown on the left side of Fig.~\ref{fig:asym}.
	
Let us assume that \(n_1 \leq n_2\), then we can estimate \(\left\{f\left((-n_1+\frac 12) w\right), \ldots,f\left(-\frac w 2\right),f\left(\frac  w 2\right), \ldots,f\left((n_1-\frac 1 2) w\right)\right\}\), we denote these points by the vector \(\mathbf f \equiv (f_1, f_2,\ldots, f_{2n_1})^\top\).

 Using these notations, we have the matrix equation
				\[\left(\begin{array}{c}\mathbf A\\\mathbf B\end{array}\right)\cdot\mathbf f = \left(\begin{array}{c}\mathbf g\\\mathbf h\end{array}\right),\] 
				with \(A_{ij}=\delta_{j,n_1+1-i}+\delta_{j,n_1+i}\), \(i=1,\ldots, n_1\), \(j=1,\ldots, 2n_1\); \\
				and \(B_{ij}=\delta_{j,n_1-\Delta n+i}+\delta_{j,n_1+1-\Delta n-i}\), \(\Delta n = n_2 -n_1\), \(i=1,\ldots, n_2\), \(j=1,\ldots, 2n_1\); with \(\delta_{i,j}\) denoting the Kronecker delta symbol.\\
				That is, we have \(2n_1\) unknowns and \(2n_1+\Delta n\) equations, which is overdetermined. Due to the randomness of the histogram, however, the best solution is not discarding \(\Delta n\) of these, but rather solve the equation in the least squares sense:
				\[\hat{ \mathbf f} = \mathop{\mathrm{argmin}}_{\mathbf f}\Bigg|\Bigg|\left(\begin{array}{c}\mathbf A\\\mathbf B\end{array}\right)\cdot\mathbf f - \left(\begin{array}{c}\mathbf g\\\mathbf h\end{array}\right)\Bigg|\Bigg|^2,\] 
				The result of this minimization along with the true PDF \(f(x)\) is shown on the right side of Fig.~\ref{fig:asym}. Note that even though the procedure can be refined by taking into account the errors of \(\mathbf g\) and \(\mathbf h\), that is, by solving a weighted least squares problem, we have found that this does not result in any improvement. Constraining the minimization to \(f_i>0\), however, visibly improves the PDF-estimates.
				
On the other hand, for \(n_1 > n_2\), by defining \(\tilde X \equiv -X-d\), we arrive at the previous case, that is, we can estimate \(\tilde f(x)\), from which it is easy to calculate \(f(x)\). 

Finally, we can use this method also for two different displacements $d_1$ and $d_2$: we first displace our whole coordinate system with $d_1$, apply the above described method for a single $d_2-d_1$ displacement (+the original one), and then we displace the result by $-d_1$.

\begin{figure}[!b]
\begin{center}
  \includegraphics[width=0.45\columnwidth]{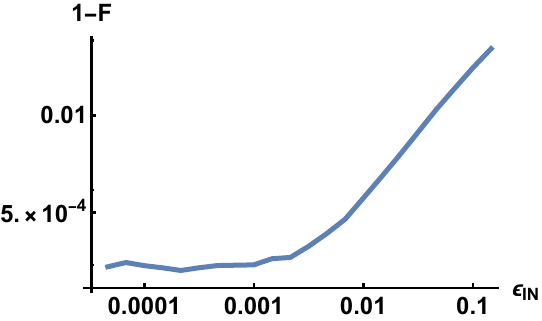}
\includegraphics[width=0.45\columnwidth]{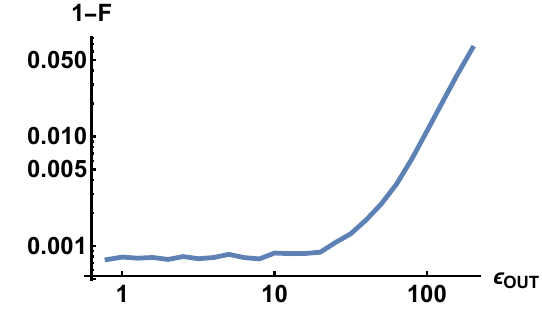}
\includegraphics[width=0.45\columnwidth]{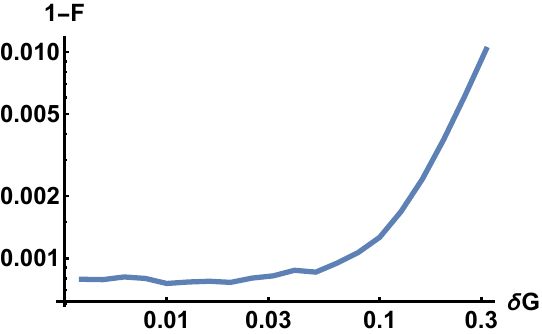}
  \includegraphics[width=0.45\columnwidth]{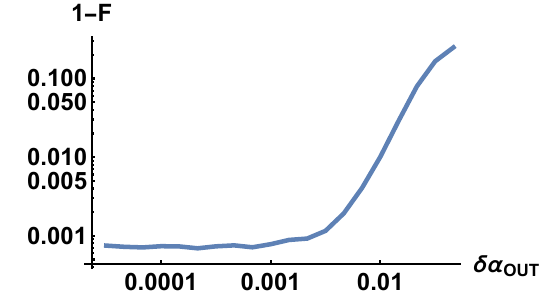}
\caption{Robustness analysis against different imperfections. Average error as a function of (top left) pre-amplification noise $\varepsilon_{\mathrm{in}}$, (top right) post-amplification noise  $\varepsilon_{out}$  (bottom left) amplification fluctuation $\delta G$, (bottom right) post-amplification loss fluctuation $\delta \alpha_{out}$.\label{fig:robust1}}
\end{center}
\end{figure}

\section{Robustness analysis} 

Checking the robustness of our method (see Fig.\ \ref{fig:robust1}), we can conclude that if the imperfections are not too large, we can achieve a similarly accurate tomography. We see quite similar behavior when looking at the fluctuations of different parameters: below a threshold, they do not affect the error of the estimated histogram, while beyond it, error starts to increase rapidly. So, keeping these parameters below these threshold values, we have reasonable room for possible uncertainties where the algorithm efficiently reconstructs the state. Note that in our simulations, by default, we used the parameters: $\varepsilon_{\mathrm{\mathrm{in}}} = 0.01$ or $0.05$ (pre-amplification noise),  $\varepsilon_{\mathrm{out}}=3$ (post-amplification noise),  $\delta G=0.01$ (standard deviation of amplification strength), $\delta\alpha_{\mathrm{out}}=10^{-3}$ (standard deviation of post-amplification transmittance), $\alpha_{\mathrm{out}} = 0.1$ (post-amplification transmittance). These figures show that the default set of parameters is well within the safe zone. The only exception to this is the pre-amplification noise $\varepsilon_{\mathrm{\mathrm{in}}}$.

From a technical perspective, it is essential to minimize pre-amplification noises, as that is the only case where we cannot easily achieve the high-accuracy regime. Also, note that, in essence, our algorithm estimates the convolution of the pre-amplification noise and the actual state of interest. This noise is amplified to the same extent as the signal, so this problem cannot be circumvented by other improvements (e.g., by increasing the value of OPA amplification $G$). So when the value of $\varepsilon_{\mathrm{in}}$ is large, the estimated histogram will not have a small error. Knowing its value, it is possible to deconvolute it from the histogram. However, this is a separate task and does not work as well if the state is not close to Gaussian. Therefore, we have omitted the deconvolution problem from our discussions.

\begin{figure}[htbp]
\begin{center}
  \includegraphics[width=0.45\columnwidth]{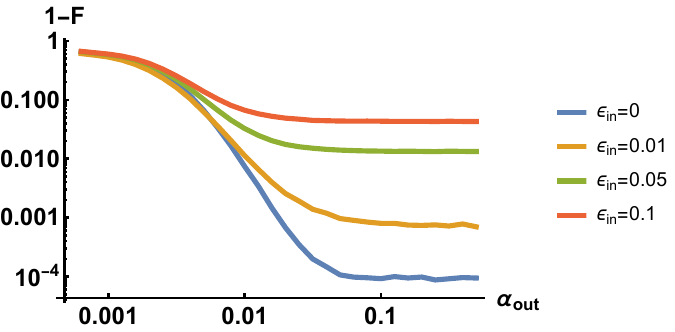}
\includegraphics[width=0.45\columnwidth]{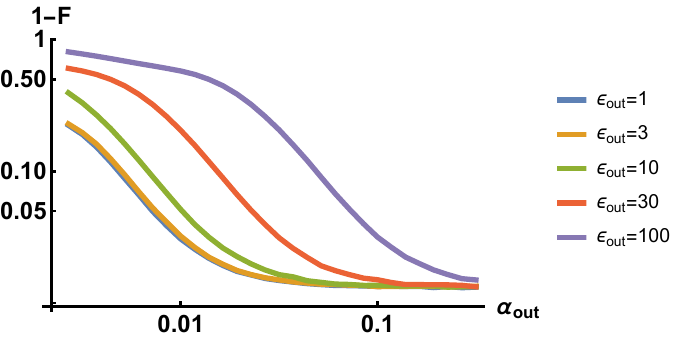}
\includegraphics[width=0.45\columnwidth]{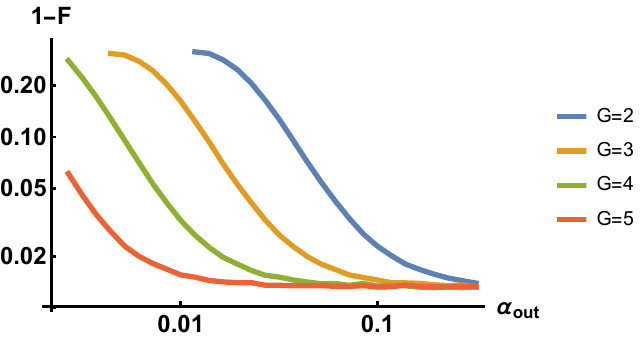}
  \includegraphics[width=0.45\columnwidth]{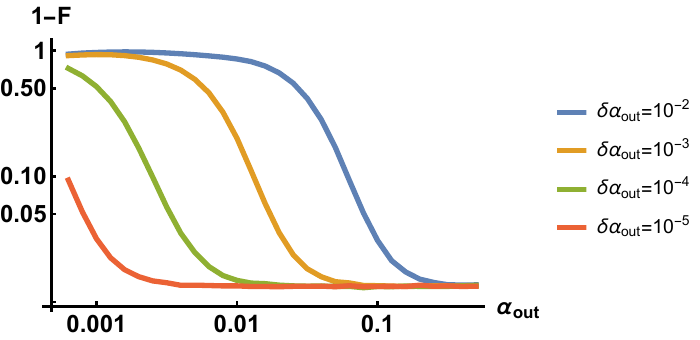}
 \caption{Analysis of maximum tolerable post-amplification losses. Average error plotted for different values of (top left) pre-amplification noise $\varepsilon_{\mathrm{in}}$, (top right) post-amplification noise  $\varepsilon_{out}$  (bottom left) amplification strength $ G$, (bottom right) post-amplification loss fluctuation $\delta \alpha_{out}$.
 \label{fig:robust2}}
\end{center}
\end{figure}

Another important observation is that the given parameters of the system are strongly connected to the performance of estimation. In Fig.\ \ref{fig:robust2}, we show this effect by plotting the error as a function of post-amplification transmittance $\alpha_{out}$. We can see that if we change the value of another parameter, there is always a visible change in the performance. When increasing pre-amplification noise (top left subfigure), the saturated level of error also increases significantly (for the reason discussed earlier). But in the case of other parameters (i.e., for post-amplification noise  $\varepsilon_{out}$, amplification strength $ G$, and post-amplification loss fluctuation  $\delta \alpha_{out}$), the observed error is only different for relatively low values of post-amplification transmittance. If the loss is low enough (i.e., the transmittance $\alpha_{out}$ is large), the error saturates at the same optimal level. This means that except for the pre-amplification noise, we can compensate for all other imperfections by improving another parameter. In our example, to achieve optimal performance we decreased the post-amplification loss sufficiently, but it works similarly for other parameters (like increasing the OPA amplification $G$). 

\begin{figure}[htbp]
\begin{center}
	\includegraphics[width=0.45\columnwidth]{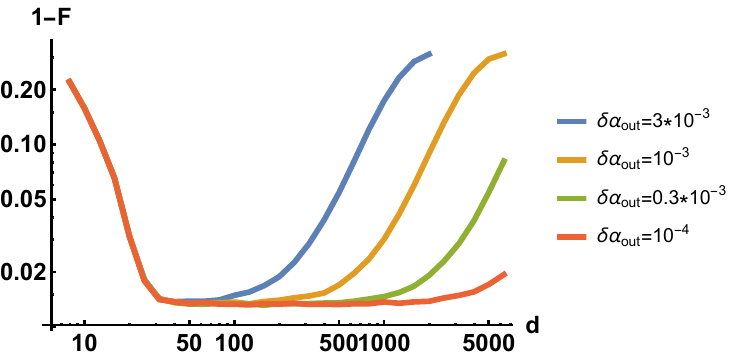}
\includegraphics[width=0.45\columnwidth]{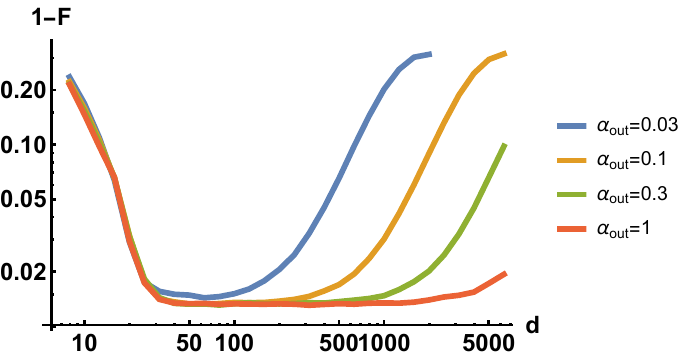}
 \caption{Analysis of the region of optimal displacement values $d$. Average error plotted for different values of (left) post-amplification loss fluctuation $\delta \alpha_{out}$ (right) post-amplification loss $\alpha_{out}$.\label{fig:robust3}}
\end{center}
\end{figure}

Finally, let us address the comment from the main text that for too large values of $d$, the estimation becomes worse and worse. This is because the effect of any type of fluctuation increases for large values of the displacement $d$. In our model, this is addressed in the parameter of post-amplification loss fluctuation $\delta \alpha_{out}$. From Fig.\ \ref{fig:robust3} left subfigure, we can see that by decreasing this fluctuation, the optimal region of displacement $d$ increases, i.e., we could use even larger values of displacement without issues. Note that we do not know the magnitude of the actual fluctuation of loss $\delta \alpha_{out}$ in experiments, but it is considered to be sufficiently low, so in practice, we expect a relatively broad interval of displacements $d$ to produce the same optimal performance. Note that a similar effect can be achieved by decreasing the value of post-amplification loss (Fig.\ \ref{fig:robust3}, right subfigure). Nevertheless, even for a 90\% loss ($\alpha_{out}=0.1$), optimal performance is attainable for a wide range of values fo the displacement $d$.

\end{document}